\documentclass{article}

\usepackage{arxiv}
\usepackage{amsmath}
\usepackage{graphicx}
\usepackage[utf8]{inputenc} 
\usepackage[T1]{fontenc}    
\usepackage{hyperref}       
\usepackage{url}            
\usepackage{booktabs}       
\usepackage{amsfonts}       
\usepackage{nicefrac}       
\usepackage{microtype}      
\usepackage{lipsum}
\usepackage{axodraw2}
\usepackage{braket}
\usepackage{slashed}

\title{New Results for the Five Point Function}

\author{
Y. Liu \\
Baylor University, Waco, TX, USA\\
\texttt{yang\_liu1@baylor.edu}\\
\And 
B.F.L. Ward \\
Baylor University, Waco, TX, USA\\
\texttt{bfl\_ward@baylor.edu}}

\begin{document}
\maketitle

\begin{abstract}
We present the new results of the computed general five-point functionby the magic spinor product method in loop integrals proposed by one (BFLW) of us. The result from the magic spinor product method agrees with that from LoopTools overall. Such encouraging agreements indicate that the magic spinor product method is a reliable approach with good efficiency and numerical stability.
\vskip .25in
\centerline{BU-HEPP-21-02}
\end{abstract}


\section{Introduction}

The one loop integrals play an important role in calculating radiative corrections in particle physics since they are manageable for fast MC event generator execution for arbitrary masses and kinematics for high energy scattering processes. It has been demonstrated that $n$-point functions at one-loop level can be reduced to scalar one loop functions \cite{Denner, PV}. Considering that representations of the scalar four-point function for arbitrary masses and momenta relevant to most high energy collider applications have been given and they fit MC implementation well, it is natural to attempt expressing higher point-function ($n\ge5$) in terms of the 1, 2, 3 and 4-point functions. When the problem becomes reducing higher-point functions into expressions in terms of the 1,2,3 and 4-point functions, we are most concerned about the numerical stability. To solve this problem,  one of us (BFLW) presented an approach to evaluate higher point loop integrals using Chinese magic in the virtual loop integration variable \cite{MagicSpinor}, which is called "magic spinor product methods in loop integrals". Based on this method, we developed a program to compute the general five-point function numerically, which we present in what follows.\par
The paper is organized as follows. In the next Section, we introduce the physics content of "magic spinor product methods in loop integrals". In Section 3, we compare our results with those from LoopTools \cite{LT}, which is a package for computation of one-loop integrals based on the FF package by G. J. van. Oldenborgh\cite{vOV90}. By comparison, it shows the result from our program is accurate and numerically stable. Section 4 contains our summary remarks.

\section{Magic Spinor Product Approach in Loop Integrals}
\label{sec:headings}

We will give a brief introduction of the "Magic Spinor Product Approach"\cite{MagicSpinor}. Here we use the conventions of Refs. \cite{GPS,JWZ2001prd} which are derived from Refs. \cite{ChnMag,KS}. The 1PI five-point function which we analyze is shown in diagram (c) in Fig.~1. It has many applications in collider precision physics. For example, together with diagrams (a) and (b) it generates a gauge invariant contribution to the ISR for $e^+e^-\to f\bar{f}+\gamma$, $f\neq e$. We focus on the application of Chinese magic in the loop integral in Fig.~1(c) to illustrate the possible simplifications here.

\begin{center}

	\begin{axopicture}(520,210)
		\Photon(50,170)(130,170){3}{7}
		\Photon(50,90)(130,90){3}{7}
		\Line(50,90)(50,170) 
		\Line(130,90)(130,170)
		\Line[arrow](5,215)(27.5,192.5)\Line[arrow](27.5,192.5)(50,170)\Text(0,210){$p_1$}
		\Photon(27.5,192.5)(57.5,222.5){2}{5}
		\Line[arrow](175,215)(130,170)\Text(160,210){$p_4$}
		\Line[arrow](5,45)(50,90)  \Text(0,50){$p_2$} 
		\Line[arrow](175,45)(130,90) \Text(160,50){$p_3$}
		\Text(67.5,220){$k$}
		\Text(90,180){$q$}
		\Text(90,20){(a)}
		
		\Photon(260,170)(340,170){3}{7}
		\Photon(260,90)(340,90){3}{7}
		\Line(260,90)(260,170)  
		\Line(340,90)(340,170)
		\Line[arrow](215,215)(260,170)
		\Line[arrow](385,215)(340,170)
		\Line[arrow](215,45)(237.5,67.5)\Line[arrow](237.5,67.5)(260,90)
		\Line[arrow](385,45)(340,90)
		\Text(210,50){$p_2$} 
		\Text(210,210){$p_1$}
		\Text(370,210){$p_4$}
		\Text(370,50){$p_3$}
		\Photon(237.5,67.5)(267.5,37.5){2}{5}
		\Text(270,45){$k$}
		\Text(300,180){$q$}
		\Text(300,20){(b)}
	\end{axopicture}
\end{center}

\begin{center}
	\begin{axopicture}(260,210)
		
		\Photon(50,170)(130,170){3}{7}
		\Photon(50,90)(130,90){3}{7}
		\Line(50,90)(50,130)\Line(50,130)(50,170)
		\Photon(5,130)(50,130){2}{5} \Text(0,130){$k$}
		\Line(130,90)(130,170)
		\Line[arrow](5,215)(50,170)\Text(0,210){$p_1$}
		
		\Line[arrow](175,215)(130,170)\Text(160,210){$p_4$}
		\Line[arrow](5,45)(50,90)  \Text(0,50){$p_2$} 
		\Line[arrow](175,45)(130,90) \Text(160,50){$p_3$}
		\Text(90,180){$q$}
		\Text(90,20){(c)}
		
	\end{axopicture}\end{center}
{ {\rm Figure 1:} ISR five-point function contributions with fermion and vector boson masses $m_f$, $m_B$, $f=1,2$, $B=V_1,V_2$ and with momenta $p_i$, $k$ with $Q\equiv p_1+p_2$ and $s=Q^2$. $p_1$ is the incoming fermion momentum, $p_2$ is the incoming antifermion momentum and $k$ is the real photon momentum. $q$ is a loop momentum.}

Applying the Feynman rules, we have
\begin{eqnarray}
&&M^{(1c)}_{\lambda_1\lambda_2\lambda'_1\lambda'_2\lambda_\gamma}=(2\pi)^4\delta(p_1+p_2-p'_1-p'_2-k)\mathcal{C}\nonumber\\
&&\times\int\frac{d^4q}{(2\pi)^4}\frac{\bar{v}_{\lambda_2}\gamma^\beta(\slashed q+\slashed p_1-\slashed k+m_1)\slashed \epsilon_{\lambda_\gamma}^\ast(\slashed q+\slashed p_1+m_1)\gamma^\alpha u_{\lambda 1}}{[(q+p_1-k)^2-m_1^2+i\epsilon][(q+p_1)^2-m_1^2+i\epsilon]}\nonumber\\
&&\frac{\bar{u}'_{\lambda'_1}\gamma_\alpha(\slashed q+\slashed p'_1+m_2)\gamma_\beta v'_{\lambda'_2}}{[(q+p_1+p_2-k)^2-M_{V_2}^2+i\epsilon][(q+p_{1}^{'2})^2-m_2^2+i\epsilon](q^2-M_{V_1}^2+i\epsilon)}\nonumber\\
&&+\cdots,
\end{eqnarray}
where we define the massless limit coupling factor
\begin{eqnarray}
\mathcal{C}&=&\mathcal{C}({\lambda_i},{\lambda'_j})\nonumber\\
&=&Q_1eG^2G'^2(v'_1+a'_1\lambda_2)(v_1-a_1\lambda_1)(v'_2+a'_2\lambda_2)(v_2+a_2\lambda'_1)
\end{eqnarray}	
with the couplings $Q_1e$, $G$, and $G'$ for the $\gamma$, $V_1$ and $V_2$ respectively. In order to obtain the loop integral in terms of Chinese magic, we take the following kinematics
\begin{align}
p_1&=(E,p\hat{z})\nonumber\\
p_2&=(E,-p\hat{z})\nonumber\\
-p_4&=(E',p'(\cos\theta'_1\hat{z}+\sin\theta'_1\hat{x}))\equiv p'_1\nonumber\\
k&=(k_0,k(\cos\theta_\gamma\hat{z}+\sin\theta_\gamma(\cos\phi_\gamma\hat{x}+\sin\phi_\gamma\hat{y})))\nonumber\\
p_1+p_2&=-p_4-p_3+k=(\sqrt{s},\vec{0})\nonumber\\
-p_3&\equiv p'_2,
\end{align}
with $k^0=k$, $\sqrt{s}=2E$. Besides, we introduce the alternative notations $p'_1=-p_4$, $p'_2=-p_3$. Now we introduce the two sets of magic polarization vectors associated to the two incoming lines
\begin{eqnarray}
(\epsilon^\mu_\sigma(\beta))^\ast&=&\frac{\bar{u}_\sigma(k)\gamma^\mu u_\sigma(\beta)}{\sqrt{2}\bar{u}_{-\sigma}u_\sigma(\beta)},\nonumber\\
(\epsilon^\mu_\sigma(\zeta))^\ast&=&\frac{\bar{u}_\sigma(k)\gamma^\mu u_\sigma(\zeta)}{\sqrt{2}\bar{u}_{-\sigma}u_\sigma(\zeta)},
\end{eqnarray}
with $\beta^2=0$ and $\zeta^2=0$. We choose the basis of the four-dimensional momentum space as follows:
\begin{align}
\ell_1&=(E,E\hat{z})\nonumber\\
\ell_2&=(E,-E\hat{z})\nonumber\\
\ell_3&=E\frac{\braket{\ell_2+|\gamma^\mu|\ell_1+}}{\sqrt{2}\braket{\ell_2-|\ell_1+}}=-\frac{E}{\sqrt{2}}(\hat{x}+i\hat{y})\nonumber\\
\ell_4&=E\frac{\braket{\ell_2-|\gamma^\mu|\ell_1-}}{\sqrt{2}\braket{\ell_2+|\ell_1-}}=\frac{E}{\sqrt{2}}(\hat{x}-i\hat{y})
\end{align}
where we use the Dirac notations in Refs. \cite{MagicSpinor,KS,CALKUL}. Note that all four of the basis four-vector are light-like so that they can attend in Chinese magic. 

We define the loop momentum as

\begin{equation}
q=\alpha_i\ell_i
\end{equation}
with summation over repeated indices. The coefficient $\alpha_i$'s are determined as
\begin{align}
\alpha_1&=\frac{q\ell_2}{2E^2}=\frac{1}{s}(D_3-D_2-s+2p_2k+M^2_{V_2}),\nonumber\\
\alpha_2&=\frac{q\ell_1}{2E^2}=\frac{1}{s}(D_1-D_0-M^2_{V_1}),\nonumber\\
\alpha_3&=\frac{q\ell_4}{E^2}=-\frac{q\ell_3^\ast}{E^2}=-\alpha_4^\ast,\nonumber\\
\alpha_4&=-\frac{i}{\sqrt{2s}}[c_jD_j+c_5M_{V_1}^2+c_6(M^2_{V_2}+2p_2k-s)+c_7(2kp_1)],
\end{align}
where we define the denominators as
\begin{align}
D_0&=q^2-M^2_{V_1}+i\epsilon\nonumber\\
D_1&=(q+p_1)^2-m_1^2+i\epsilon\nonumber\\
D_2&={q+p_1-k}^2-m^2_{1}+i\epsilon\nonumber\\
D_3&={q+p_1+p_2-k}^2-M^2_{V_2}+i\epsilon\nonumber\\
D_4&=(q-p_4)^2-m^2_2+i\epsilon
\end{align}
such that the expansion coefficients $\{c_j\}$ are
\begin{align}
c_0&=\csc\phi_\gamma\left(\frac{\csc\theta'_1e^{i\phi_\gamma}}{\beta'_1E'_1}-\frac{\csc\theta'_1e^{i\phi_\gamma}}{\beta'_1\sqrt{s}}+\frac{\csc\theta_\gamma}{\sqrt{s}}-\frac{\cot\theta'_1e^{i\phi_\gamma}-\cot\theta_\gamma}{\beta_1\sqrt{s}} \right),\nonumber\\
c_1&=\csc\phi_\gamma\left(\frac{\csc\theta'_1e^{i\phi_\gamma}}{\beta'_1\sqrt{s}}-\frac{\csc\theta'_1}{\sqrt{s}}+\frac{\cot\theta'_1e^{i\phi_\gamma}-\cot\theta_\gamma}{\beta_1\sqrt{s}}+\frac{\csc\theta_\gamma}{k^0} \right),\nonumber\\
c_2&=\csc\phi_\gamma\left(\frac{-\csc\theta'_1e^{i\phi_\gamma}}{\beta'_1\sqrt{s}}+\frac{\csc\theta_\gamma}{\sqrt{s}}+\frac{\cot\theta'_1e^{i\phi_\gamma}-\cot\theta_\gamma}{\beta_1\sqrt{s}}-\frac{\csc\theta_\gamma}{k^0} \right),\nonumber\\
c_3&=\csc\phi_\gamma\left(\frac{\csc\theta'_1e^{i\phi_\gamma}}{\beta'_1\sqrt{s}}-\frac{\csc\theta_\gamma}{\sqrt{s}}-\frac{\cot\theta'_1e^{i\phi_\gamma}-\cot\theta_\gamma}{\beta_1\sqrt{s}}\right),\nonumber\\
c_4&=-\csc\phi_\gamma\frac{\csc\theta'_1e^{i\phi_\gamma}}{\beta'_1E'_1},\nonumber\\
c_5&=\csc\phi_\gamma\left(\frac{\csc\theta'_1e^{i\phi_\gamma}}{\beta'_1E'_1}-\frac{\csc\theta'_1e^{i\phi_\gamma}}{\beta'_1\sqrt{s}}+\frac{\csc\theta_\gamma}{\sqrt{s}}-\frac{\cot\theta'_1e^{i\phi_\gamma}-\cot\theta_\gamma}{\beta_1\sqrt{s}} \right),\nonumber\\
c_6&=\csc\phi_\gamma\left(\frac{\csc\theta'_1e^{i\phi_\gamma}}{\beta'_1\sqrt{s}}-\frac{\csc\theta_\gamma}{\sqrt{s}}-\frac{\cot\theta'_1e^{i\phi_\gamma}-\cot\theta_\gamma}{\beta_1\sqrt{s}}\right),\nonumber\\
c_7&=-\csc\phi_\gamma\frac{\csc\theta_\gamma}{k^0},
\end{align}
with $\beta=\frac{p}{E}$ and $\beta'_1=\frac{p'}{E'}$. Therefore the $\{c_j\}$ are determined by the kinematics that we choose. Note that the Chinese magic now carries over to the loop variable via the identity
\begin{align}
\slashed q&=\alpha_j\slashed \ell_j\nonumber\\
&=\sum_{j=1}^{2}\alpha_j(\ket{\ell_j+}\bra{\ell_j+}+\ket{\ell_j-}\bra{\ell_j-})+\alpha_3\frac{\sqrt{2}E}{\braket{p_2-|p_1+}}(\ket{\ell_2-}\bra{\ell_1-}+\ket{\ell_1-}\bra{\ell_2+})\nonumber\\
&+\alpha_4\frac{\sqrt{2}E}{\braket{p_2+|p_1-}}(\ket{\ell_2+}\bra{\ell_1+}+\ket{\ell_1-}\bra{\ell_2-})\nonumber\\
&\equiv\sum_{j=1}^{2}\alpha_j(\ket{p_j+}\bra{p_j+}+\ket{p_j-}\bra{p_j-})+\alpha_3\frac{\sqrt{2}E}{\braket{p_2-|p_1+}}(\ket{p_2-}\bra{p_1-}+\ket{p_1-}\bra{p_2+})\nonumber\\
&+\alpha_4\frac{\sqrt{2}E}{\braket{p_2+|p_1-}}(\ket{p_2+}\bra{p_1+}+\ket{p_1-}\bra{p_2-})\nonumber\\
&\equiv\sum_{j=1}^{2}\alpha_j(\ket{p_j+}\bra{p_j+}+\ket{p_j-}\bra{p_j-})+\widetilde{\alpha}_3(\ket{p_2-}\bra{p_1-}+\ket{p_1-}\bra{p_2+})\nonumber\\
&+\widetilde{\alpha}_4(\ket{p_2+}\bra{p_1+}+\ket{p_1-}\bra{p_2-}),
\end{align}
where we let $\ell_1\equiv p_1$, $\ell_2\equiv p_2$ since for the numerator algebra we work in the massless limit. And we define as well
\begin{align}
\widetilde{\alpha}_3&\equiv\alpha_3\frac{\sqrt{2}E}{\braket{p_2-|p_1+}}=-\frac{\alpha_3}{\sqrt{2}},\nonumber\\
\widetilde{\alpha}_4&\equiv\alpha_4\frac{\sqrt{2}E}{\braket{p_2+|p_1-}}=-\frac{\alpha_4}{\sqrt{2}}.
\end{align}

Next, we introduce the representation (10) into the numerator, $N$, of the integrand in eq (1). Then we have the reduction
\begin{align}
N&=\frac{4\sqrt{2}}{\braket{k-|p_1+}}\{ (A_1\braket{p_2+|p'_1-}\braket{p'_2-|p_2+}\nonumber\\
& +A_2\braket{p_2+|p'_1-}\braket{p'_2-|p_1+} )\times(A_3\braket{p_2+|p'_1-}\braket{p'_1-|p_1+}\nonumber\\
& +A_4\braket{p_1+|p'_1-}\braket{p'_1-|p_1+} ) \nonumber\\
&+\widetilde{\alpha}_4 (A_1\braket{p_2+|p_1-}\braket{p'_2-|p_2+} +A_2\braket{p_2+|p_1-}\braket{p'_2-|p_1+} )
\nonumber\\
&\times(A_3\braket{p_2+|p'_1-}\braket{p_2-|p_1+}\nonumber\\
& +A_4\braket{p_1+|p'_1-}\braket{p_2-|p_1+} ) \}
\end{align}
from the standard identities
\begin{align}
\slashed \epsilon_{\lambda_\gamma}^\ast&=\frac{\sqrt{2}}{\braket{k-\lambda_\gamma|\ell_1\lambda_\gamma}}[\ket{\ell_1\lambda_\gamma}\bra{k\lambda_\gamma}+\ket{k-\lambda_\gamma}\bra{\ell_1-\lambda_\gamma}],\nonumber\\
\gamma^\rho\braket{\ell_1\lambda|\gamma_\rho|\ell_2\lambda}&=2[\ket{\ell_1-\lambda}\bra{\ell_2-\lambda}+\ket{\ell_2\lambda}\bra{\ell_1\lambda}],\nonumber\\
\slashed\ell_1&=\ket{\ell_1+}\bra{\ell_1+}+\ket{\ell_1-}\bra{\ell_1-},
\end{align}
where we defined
\begin{equation}
\begin{cases}
A_1=\widetilde{\alpha}_4\braket{p_1+|k-}+\alpha_2\braket{p_2+|k-},
\\
A_2=(1+\alpha_1)\braket{p_1+|k-}+\widetilde{\alpha}_3\braket{p_2+|k-},\\
A_3=\alpha_2\braket{p_1-|p_2+},\\
A_4=\widetilde{\alpha}_4\braket{p_1-|p_2+},
\end{cases}
\end{equation}
for the magic choice $\beta=p_1$. Note that the Chinese magic trick has eliminated all but one set of the terms with three factors of the virtual momentum expansion coefficients. Moreover, in the numerator of the propagator before or after the real emission vertex, this trick has annihilated the terms associated with $\slashed p_1\slashed k$ as well as half of the terms in the respective virtual momentum expansion in the former case. Compared to the traditional approach of taking traces on the fermion lines, the Chinese magic trick has eliminated a large fraction of the terms on the right-hand side of eq. (12).

So far we have simplified considerably the computation and have removed the Gram determinant-type factors in the tensor integral reductions. But our calculation sill depends on the Gram determinant-type denominator factors in the results in Refs. \cite{tHscarlar,WJ,DD,DD1} for the representation of the five-point function in terms of scalar four-point functions. In order to achieve a better numerical stability, we need to replace the representation of the scalar five-point function.

Let us begin with the identity
\begin{align}
q^2&=D_0+M_{V_1}^2-i\epsilon=(\alpha_i\ell_i)^2\nonumber\\
&=2\alpha_1\alpha_2\ell_1\ell_2+2\alpha_3\alpha_4\ell_3\ell_4=s\alpha_1\alpha_2+\frac{s}{2}\alpha_3\alpha_4.
\end{align}

Dividing by $D_0\cdots D_4$ and integrating over $d^4q$ we have the following representation of the required scalar five-point function
\begin{align}
&E_0(\bar{p}_1,\bar{p}_2,\bar{p}_3,\bar{p}_4,\bar{m}_0,\bar{m}_1,\bar{m}_2,\bar{m}_3,\bar{m}_4)\nonumber\\
&=\frac{1}{C_{E_0}}\biggl\{-D_0(0)+\frac{1+\beta_1^2}{2s\beta_1^2}\bigg[C_0(13)-C_0(12)-C_0(03)+C_0(02)\nonumber\\
&+(M_{V_2}^2-s+2p_2k) 
\bigg(D_0(1)-D_0(0)\bigg)-M^2_{V_1}\bigg(D_0(3)-D_0(2)\bigg)    \bigg]   \nonumber\\
&-\frac{1-\beta_1^2}{4s\beta_1^2}\bigg[\Delta r_{1,0}\bigg(D_0(1)-D_0(0)\bigg)+2\Delta\bar{p}_{1,0}\bigg( D_{11}(1)\bar{\bar{p}}(1)_1-D_{11}(0)\bar{\bar{p}}(0)_1\nonumber\\&+D_{12}(1)\bar{\bar{p}}(1)_2-D_{12}(0)\bar{\bar{p}}(0)_2+D_{13}(1)\bar{\bar{p}}(1)_3-D_{13}(2)\bar{\bar{p}}(2)_3-D_{0}(3)\bar{\bar{p}}(3)_4\nonumber\\
& +D_{0}(2)\bar{\bar{p}}(2)_4 \bigg) +2(M_{V_2}^2-s+2p_2k)\bigg(D_0(3)-D_0(2)\bigg) \bigg]  \nonumber\\
&-\frac{1}{4}\bigg[\sum_{j=0}^{4}|c_j|^2\bigg(C_0(j,j+1)+\Delta r_{j,j+1}D_0(j)+2\Delta\bar{p}_{j,j+1} (D_{11}(j)\bar{\bar{p}}(j)_1 \nonumber\\
&+D_{12}(j)\bar{\bar{p}}(j)_2+D_{13}(j)\bar{\bar{p}}(j)_3-D_{0}(j)\bar{\bar{p}}(j)_4   ) \bigg)  \nonumber\\
&+2\bigg( \sum_{i>j}^{4}\Re(c_ic^\ast_j)C_0(ij)+\sum_{j=0}^{4}\Re(c_j(c_5^\ast M_{V_1}^2+c_6^\ast(M^2_{V_2}-s+2p_2k)+c_7^\ast(2kp_1))  )\nonumber\\
&D_0(j) \bigg)\bigg]\biggr\}
\end{align}
where
$\bar{p}=p_1$, $\bar{p}_2=p_1-k$, $\bar{p}_3=p_1+p_2-k$, $\bar{p}=p'_1$, $\bar{m_0}-M_{V_1}$, $\bar{m}_1=m_1$, $\bar{m}=m_1$, $\bar{m}_3=M_{V_2}$, $\bar{m}_4=m_2$,
where the coefficient $C_{E_0}$ is 
\begin{align}
C_{E_0}=&M^2_{V_1}-i\epsilon+\frac{1+\beta_1^2}{2\beta_1^2s}M^2_{V_1}(M^2_{V_2}-s+2p_2k)+\frac{1-\beta_1^2}{4\beta_1^2s}(M^4_{V_1} \nonumber\\
&+(M^2_{V_2}-s+2p_2k)^2)+\frac{1}{2}\Re[c_5c_6^\ast M^2_{V_1}(M_{V_2}^2-s+2p_2k)  \nonumber\\
&+c_5c_7^\ast M^2_{V_1}(2kp_1)  +c_6c_7^\ast(M^2_{V_2}-s+2p_2k)(2kp_1)  ]\nonumber\\
&+\frac{1}{4}[|c_5|^2M^4_{V_1}+|c_6|^2(M_{V_2}^2-s+2p_2k)^2+|c_7|^2(2kp_1)^2].
\end{align}
We have used a mixture of notations from \cite{PV,WJ,DD,DD1} such that
\begin{align}
D_j&=(q-\bar{p}_j)^2-\bar{m}^2_j+i\epsilon=q^2+2q\bar{p}_j+\bar{p}^2_j-\bar{m}^2_j+i\epsilon\equiv q^2+2q\bar{p}_j+r_j,\nonumber\\
\Delta_{i,j}&\equiv r_i-r_j,\nonumber\\
\Delta\bar{p}_{i,j}&\equiv\bar{p}_i-\bar{p}_j\nonumber\\
D_0(j)&\equiv\text{four-point scalar function obtained from five-point scalar function}\nonumber\\
&\text{by omitting denominator} D_j,\nonumber\\
C_0(i,j)&\equiv\text{three-point scalar function obtained from five-point scalar function}\nonumber\\
&\text{by omitting denominator} D_i \text{ and } D_j, i\neq j \nonumber\\
\end{align}
where we also use the Passarino-Veltman \cite{PV} notation of the four-point one-tensor integral, $D_\mu(j)$, obtained from the five-point one tensor integral function by omitting the denominator $D_j$, with
$$D_\mu(j)\equiv D_{11}(j)\bar{\bar{p}}(j)_1+D_{12}(j)\bar{\bar{p}}(j)_2+D_{13}(j)\bar{\bar{p}}(j)_3-D_{0}(j)\bar{\bar{p}}(j)_4,
$$
where the four-vectors $\{\bar{\bar{p}}(j)_j\}$ are determined according to Ref. \cite{PV}. Note that $\bar{\bar{p}}(j)_4$ is only nonzero if it is necessary to shift the $q$-integration variable by it to reach the standard form of Passarino-Veltman representation. 

In sum, we have exhibited that the application of Chinese magic technique in the virtual loop momentum could reduce significantly the volume of algebra required for efficient and stable physical calculations of higher point virtual corrections with the general mass scale. Furthermore, we could construct computer realizations of the method described above for evaluating the five-point function. In the next section, we will exhibit our results and compare them with those from LoopTools.

\section{Numerical Results for the Five-Point Function $E_0$}
Based on the magic spinor product method, we have developed a computer program to calculate the 1PI scalar five-point function $E_0$. In our program, the LoopTools package is included. The 1PI five-point function $E_0,$ which results from the graph in Fig. 2 when all lines are scalars and have masses as indicated, is calculated with the help of eqs. (9), (16) and (17). Scalar three-point functions $C_0(i,j)$, scalar four-point functions $D_0(j)$ and tensor four-point functions $D_{\mu\nu}(j)$ are calculated by Looptools. To be specific, we choose, using a standard notation,
\begin{equation*}
\begin{cases}
\sqrt{s}=500\text{ GeV},\\
m_1=m_2=m_e=0.510999\times 10^{-3}\text{ GeV},\\
m_5=m_\mu=0.1056583\text{ GeV},\\
M_{V_1}=M_{V_2}=91\text{ GeV}\cong m_Z,
\end{cases}
\end{equation*}
and the kinematics is determined by eq. (3).

\begin{center}
	\begin{axopicture}(210,200)
		\Photon(60,140)(60,60){3}{6}
		\Line(60,140)(100,140)\Line(100,140)(140,140)
		\Photon(140,140)(140,60){3}{6}
		\Line(60,60)(140,60)
		\Line[arrow](15,185)(60,140)
		\Line[arrow](185,185)(140,140)
		\Line[arrow](15,15)(60,60)
		\Line[arrow](185,15)(140,60)
		\Photon(100,140)(100,190){3}{5}
		\Text(10,190){$p_1$}
		\Text(190,190){$p_2$}
		\Text(190,10){$p_3$}
		\Text(10,10){$p_4$}
		\Text(100,200){$k$}
		\Text(80,145){$m_1$}
		\Text(120,145){$m_2$}
		\Text(45,100){$M_{V_1}$}
		\Text(155,100){$M_{V_2}$}
		\Text(100,50){$m_5$}
	\end{axopicture}
\end{center}
{{\rm Figure 2:} ISR 1PI five-point function with fermion and vector boson masses $m_1=m_2=m_e$,$\; m_5=m_\mu $, $M_{V_1}=M_{V_2}=m_Z$, and with momenta $p_i$, $k$ as given in Fig.~1.}

then by eq. (8) and (18), we have the expressions of $D_0(j)$ and $C_0(i,j)$ which give inputs of the corresponding three- and four-point functions in Looptools so that we can compute $D_0(j)$, $C_0(i,j)$ and $D_{\mu\nu}(i,j)$ via Looptools. Besides, we have computed $E_0$ via Looptools directly as a comparison. We choose $(\phi_\gamma,\theta_\gamma,\theta'_1)$ as the variables. The percentage differences of $E_0(\theta_\gamma,\phi_\gamma)$ and $E_0^{LoopTools}(\theta_\gamma,\phi_\gamma)$ are shown in Figs.~ 3 -- 15.

\setcounter{figure}{2}

\begin{figure}
	\begin{center}
		\includegraphics[scale=0.55]{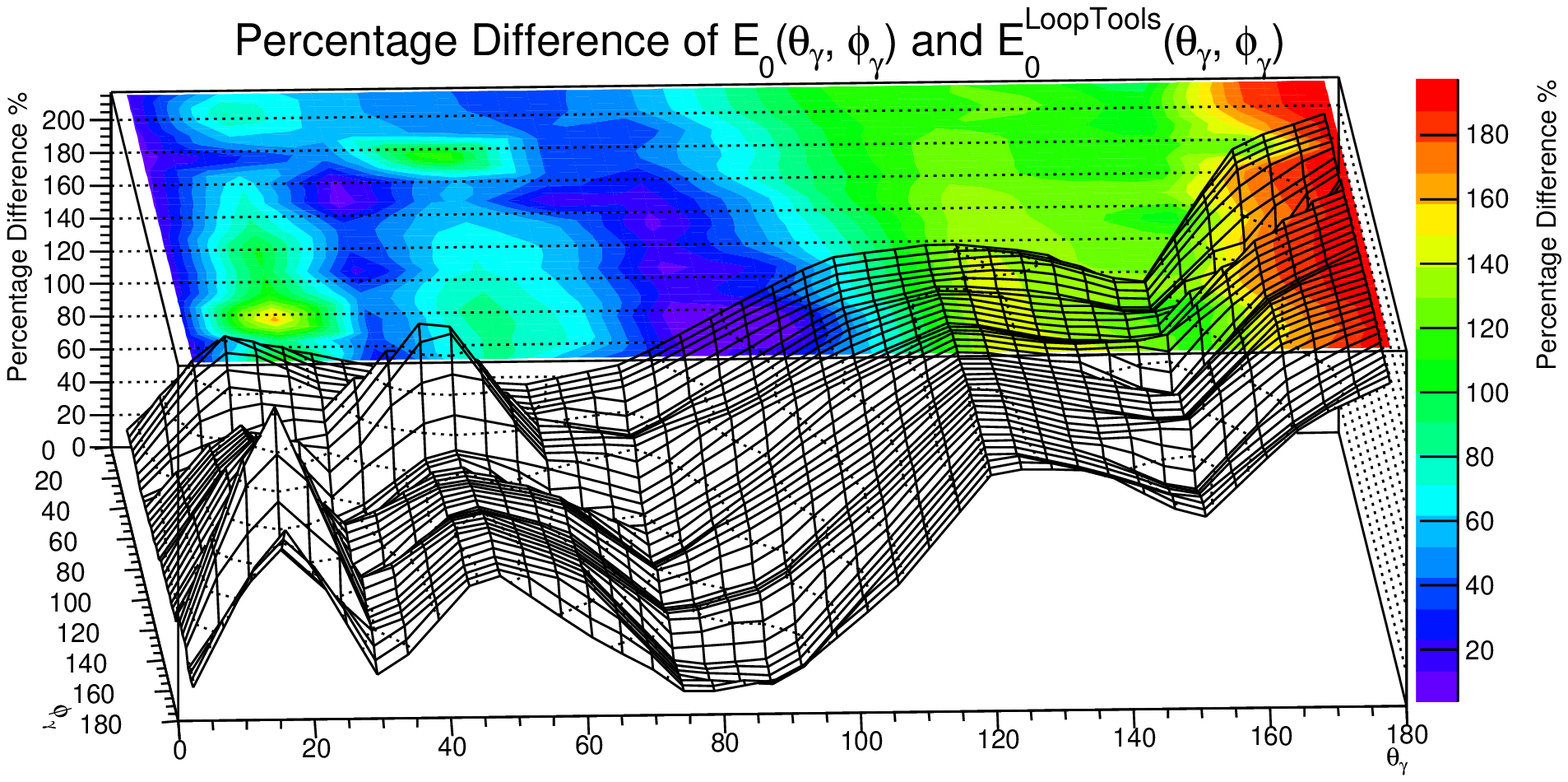}
		\caption{ Percentage difference of $E_0(\theta_\gamma,\phi_\gamma)$ and $E_0^{LoopTools}(\theta_\gamma,\phi_\gamma)$ for $\sqrt{s}=500\text{ GeV}$, $M_{V_1}=91\text{ GeV}$, $M_{V_1}=91\text{ GeV}$ and $\theta'_1=0^{\circ}$ }
	\end{center}
\end{figure}
In Fig.~3, we see that there is not much variation with $\phi_\gamma$ and that, near $\theta_\gamma = 0$, which is the regime that is collinear with the incoming $e^-$ and the outgoing $\mu^-$, we have very close agreement between the two results but that, as the $\theta_\gamma$ approaches $180^{\circ}$, the two results differ but they are still within a factor of $\sim 2$ of one another.

\begin{figure}
	\begin{center}
		\includegraphics[scale=0.55]{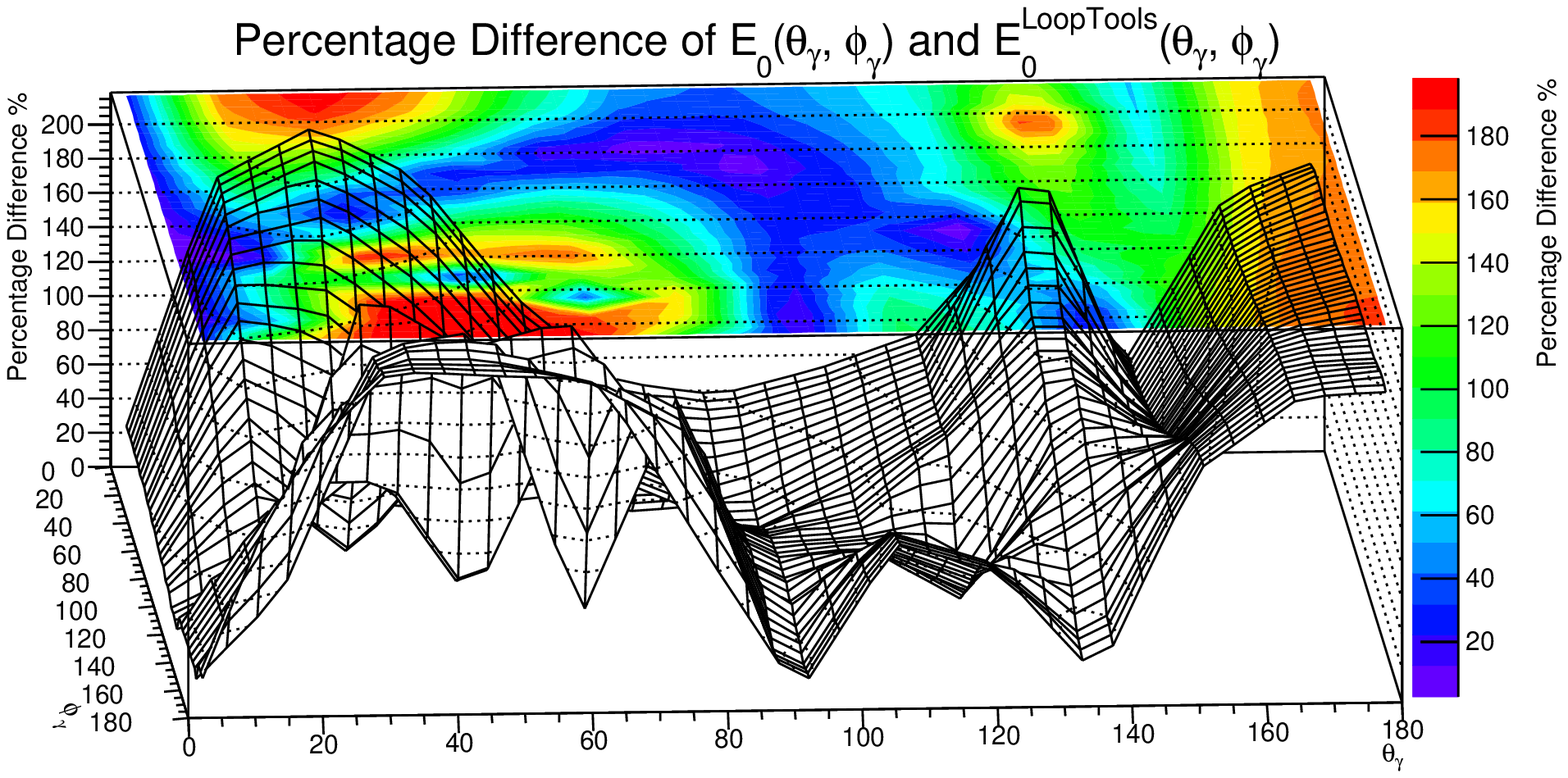}
		\caption{ Percentage difference of $E_0(\theta_\gamma,\phi_\gamma)$ and $E_0^{LoopTools}(\theta_\gamma,\phi_\gamma)$ for $\sqrt{s}=500\text{ GeV}$, $M_{V_1}=91\text{ GeV}$, $M_{V_1}=91\text{ GeV}$ and $\theta'_1=15^{\circ}$ }
	\end{center}
\end{figure}
In Fig.~4, where the scattering angle between the outgoing $\mu^-$ and the incoming $e^-$ is $\theta'_1 = 15^{\circ}$, near the incoming $e^-$ we see close agreement and small variation with $\phi_\gamma$. As we approach the other collinear regions, we see less agreement, but again the two results are within a factor of $\sim 2$ of one anoher.

\begin{figure}
	\begin{center}
		\includegraphics[scale=0.55]{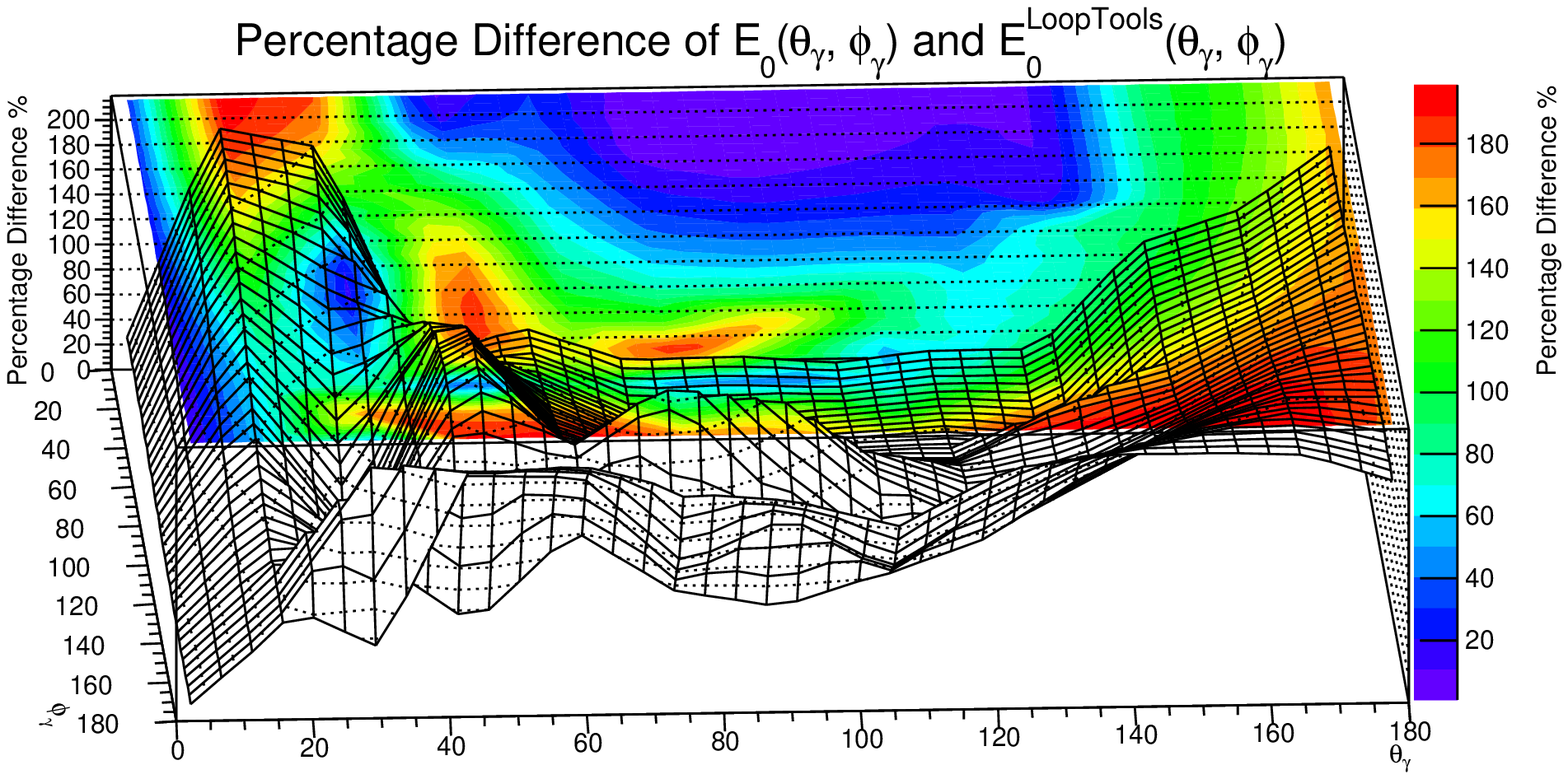}\\
		\caption{ Percentage difference of $E_0(\theta_\gamma,\phi_\gamma)$ and $E_0^{LoopTools}(\theta_\gamma,\phi_\gamma)$ for $\sqrt{s}=500\text{ GeV}$, $M_{V_1}=91\text{ GeV}$, $M_{V_1}=91\text{ GeV}$ and $\theta'_1=30^{\circ}$ }
	\end{center}
\end{figure}
In Fig.~5, where $\theta'_1 = 30^{\circ}$, we again see close agreement near the incoming $e^-$ with small variation with $\phi_\gamma$. As one approaches the other charged partciles one again sees less agreement but the percentage of the phase space wherein this occurs is somewhat less than that in Fig.~4, for comparison. The two results are within a factor $\sim 2$ of each other.

\begin{figure}
	\begin{center}
		\includegraphics[scale=0.55]{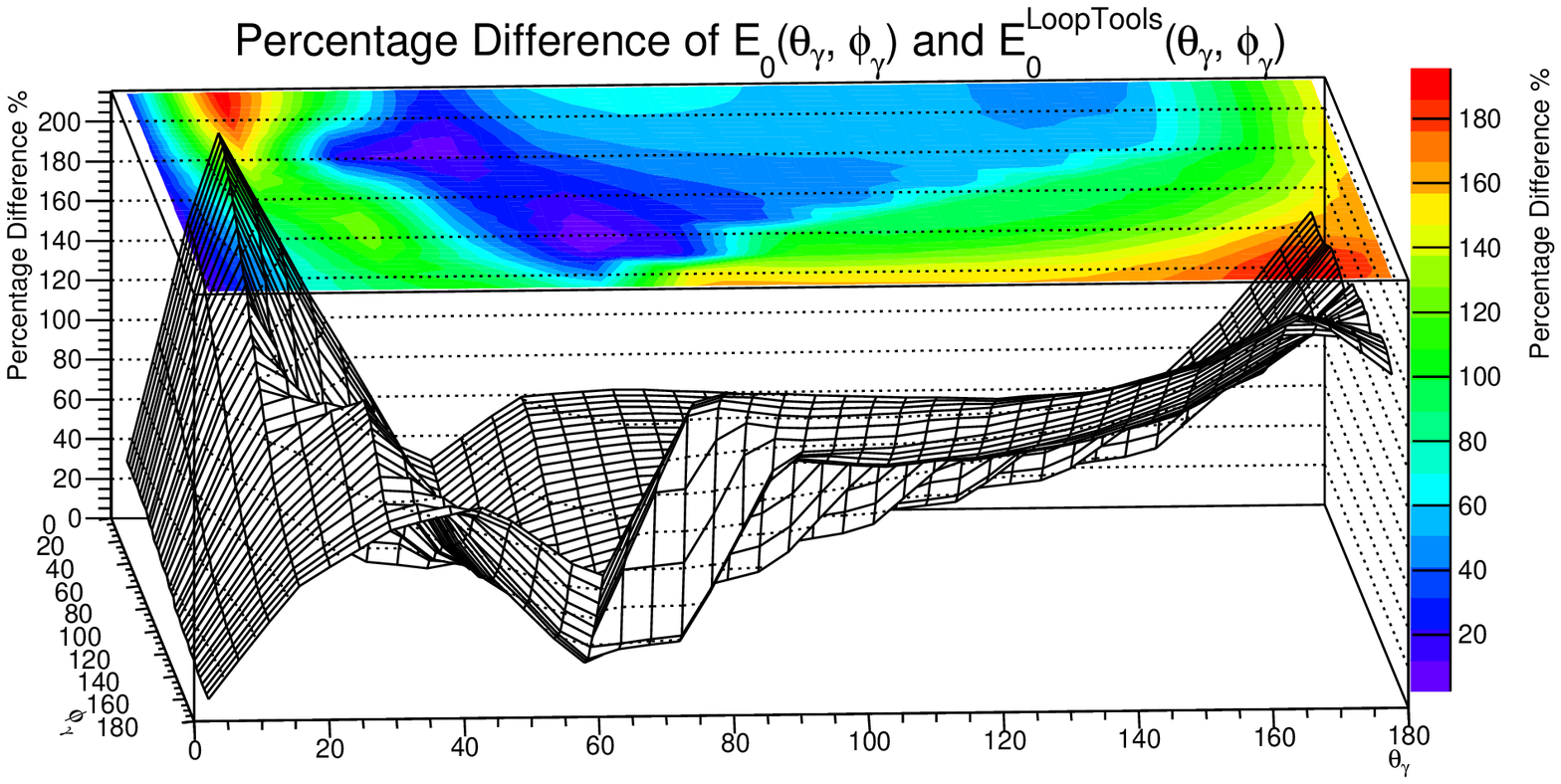}\\
		\caption{ Percentage difference of $E_0(\theta_\gamma,\phi_\gamma)$ and $E_0^{LoopTools}(\theta_\gamma,\phi_\gamma)$ for $\sqrt{s}=500\text{ GeV}$, $M_{V_1}=91\text{ GeV}$, $M_{V_1}=91\text{ GeV}$ and $\theta'_1=45^{\circ}$ }
	\end{center}
\end{figure}
In Fig.~6, where $\theta'_1 = 45^{\circ}$, the trend seen in Figs.~3 - 5 continues,
with the added feature that the region wherein one has greater than 100\% differences now significantly reduced. The two results are within a factor $\sim 2$ of one another.

\begin{figure}
	\begin{center}
		\includegraphics[scale=0.55]{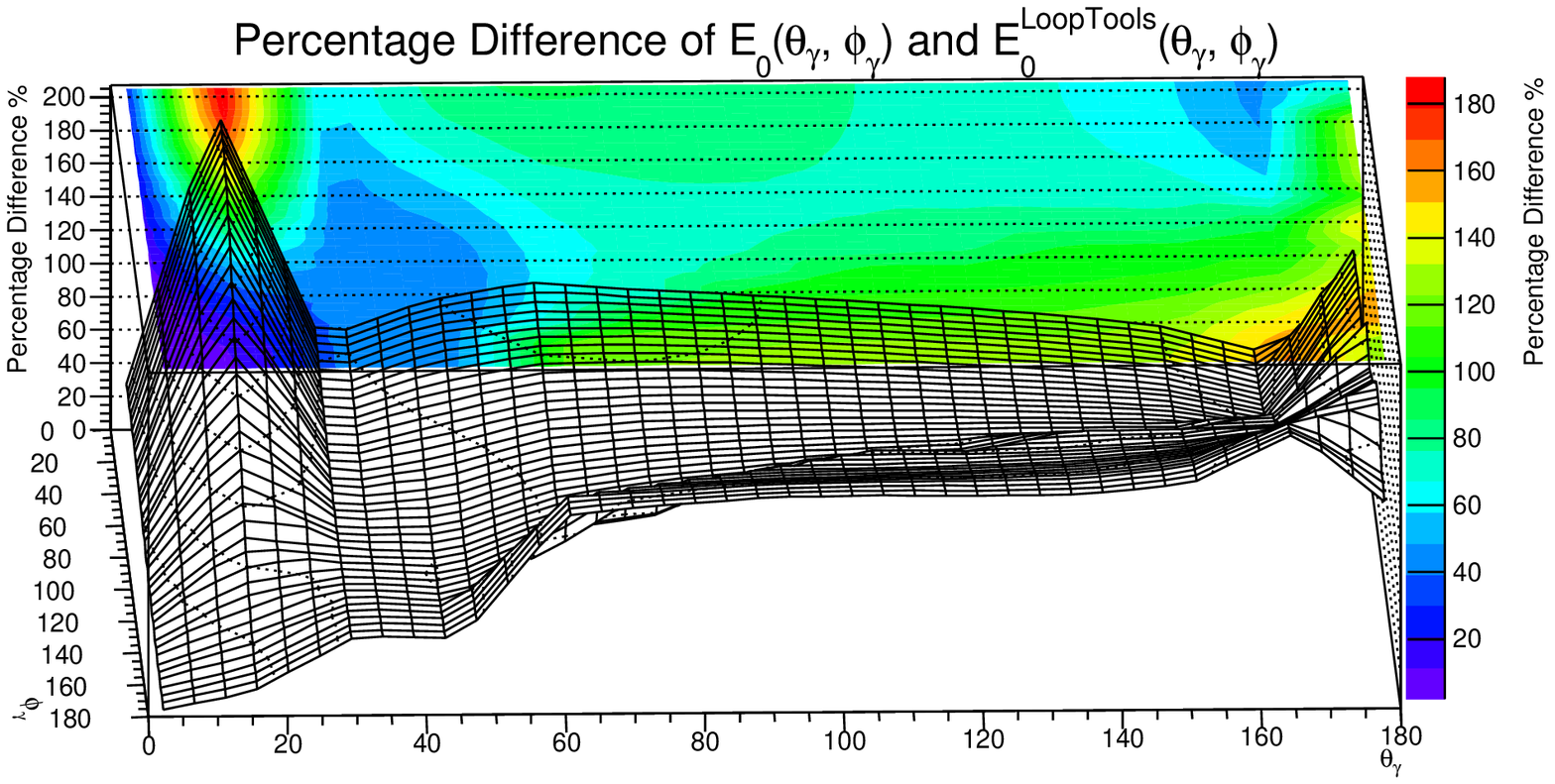}\\
		\caption{ Percentage difference of $E_0(\theta_\gamma,\phi_\gamma)$ and $E_0^{LoopTools}(\theta_\gamma,\phi_\gamma)$ for $\sqrt{s}=500\text{ GeV}$, $M_{V_1}=91\text{ GeV}$, $M_{V_1}=91\text{ GeV}$ and $\theta'_1=60^{\circ}$ }
	\end{center}
\end{figure}
In Fig.~7, where $\theta'_1 = 60^{\circ}$, with one sees a continuation of what one has in Fig.~6, with an even greater reduction in the size of the region in which the two results differ by more tha 100\%. Again, the two results are always within a factor $\sim 2$ of one another.

\begin{figure}
	\begin{center}
		\includegraphics[scale=0.55]{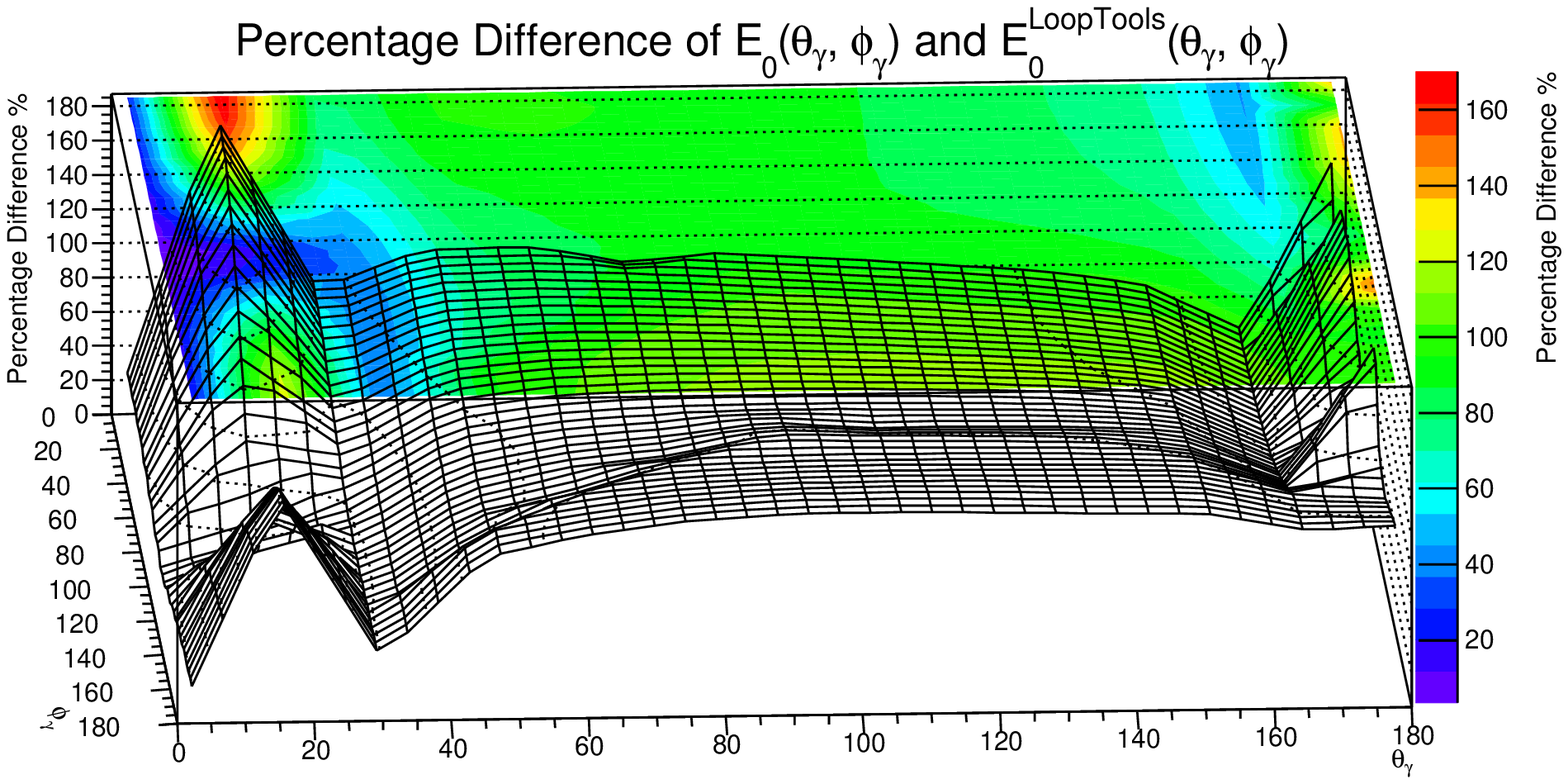}\\
		\caption{ Percentage difference of $E_0(\phi_\gamma,\theta_\gamma)$ and $E_0^{LoopTools}(\theta_\gamma,\phi_\gamma)$ for $\sqrt{s}=500\text{ GeV}$, $M_{V_1}=91\text{ GeV}$, $M_{V_1}=91\text{ GeV}$ and $\theta'_1=75^{\circ}$ }
	\end{center}
\end{figure}
In Fig.~8, where $\theta'_1 = 75^{\circ}$, there are just a small region near $(\theta_\gamma,\;\phi_\gamma) \cong (20^{\circ}, \;0^{\circ})$ and two smaller regions near the incoming $e^+$ where the two results differ by more than ~ 100\%. They remain within a factor $\sim 2$ of one another.

\begin{figure}
	\begin{center}
		\includegraphics[scale=0.55]{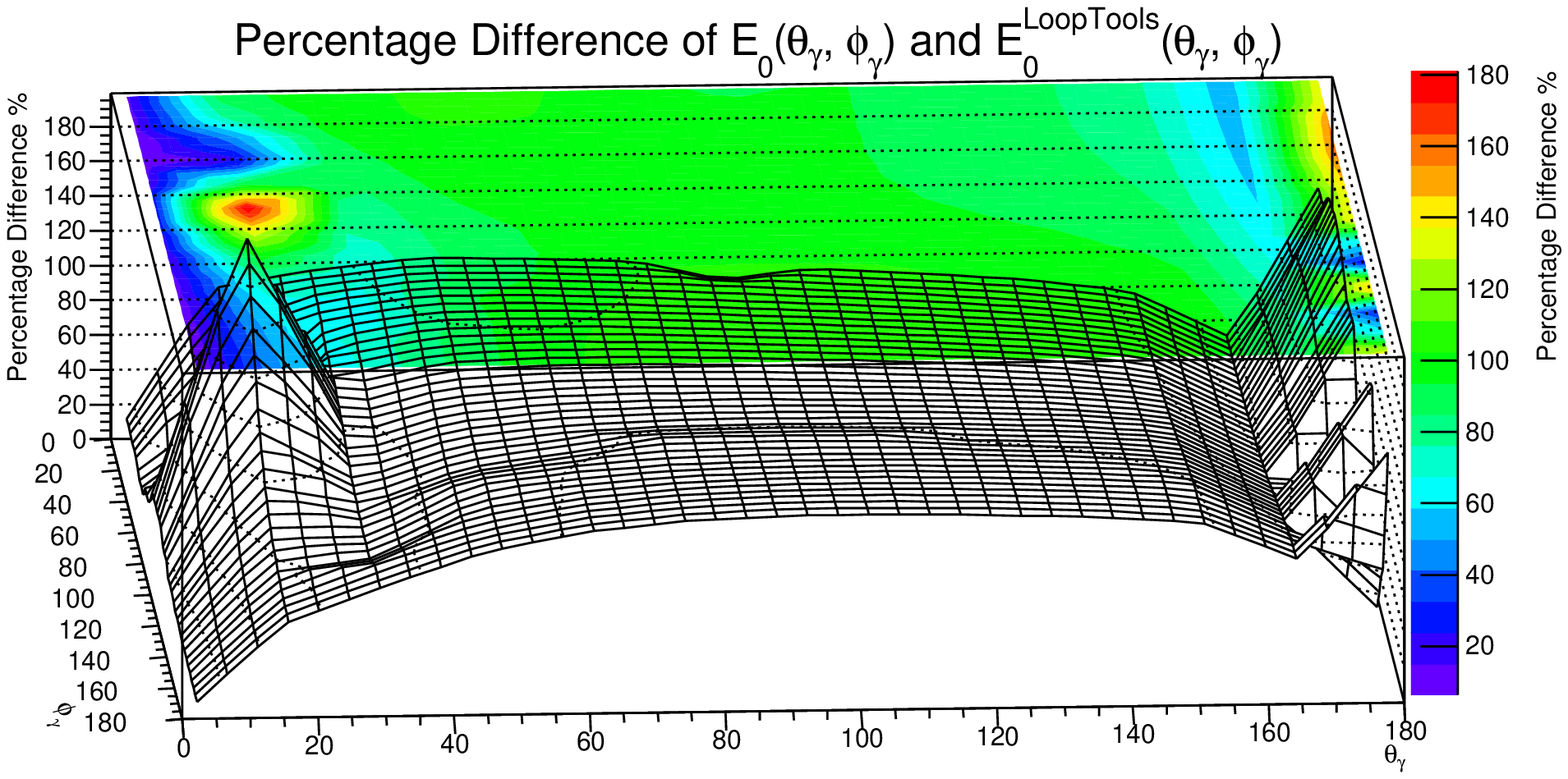}\\
		\caption{ Percentage difference of $E_0(\theta_\gamma,\phi_\gamma)$ and $E_0^{LoopTools}(\theta_\gamma,\phi_\gamma)$ for $\sqrt{s}=500\text{ GeV}$, $M_{V_1}=91\text{ GeV}$, $M_{V_1}=91\text{ GeV}$ and $\theta'_1=90^{\circ}$ }
	\end{center}
\end{figure}
In Fig.~9, where $\theta'_1 = 90^{\circ}$, there are a small region near $(\theta_\gamma,\;\phi_\gamma) \cong (20^{\circ}, 80^{\circ})$ and a small region near the incoming $e^+$ where the two results differ by more than ~ 100\%. They remain within a factor $\sim 2$ of one another.

\begin{figure}
	\begin{center}
		\includegraphics[scale=0.55]{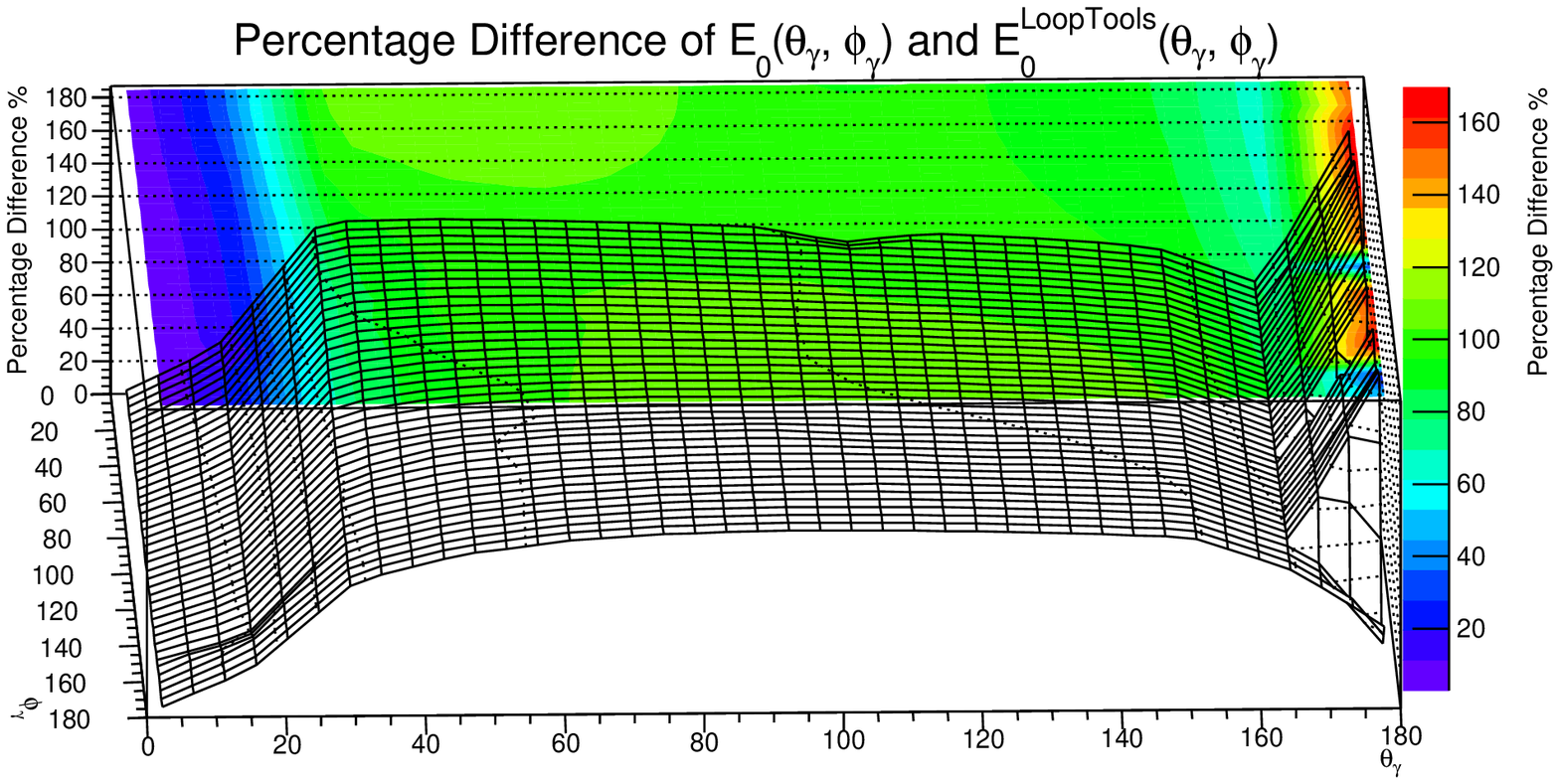}\\
		\caption{ Percentage difference of $E_0(\theta_\gamma,\phi_\gamma)$ and $E_0^{LoopTools}(\theta_\gamma,\phi_\gamma)$ for $\sqrt{s}=500\text{ GeV}$, $M_{V_1}=91\text{ GeV}$, $M_{V_1}=91\text{ GeV}$ and $\theta'_1=105^{\circ}$ }
	\end{center}
\end{figure}

\begin{figure}
	\begin{center}
		\includegraphics[scale=0.55]{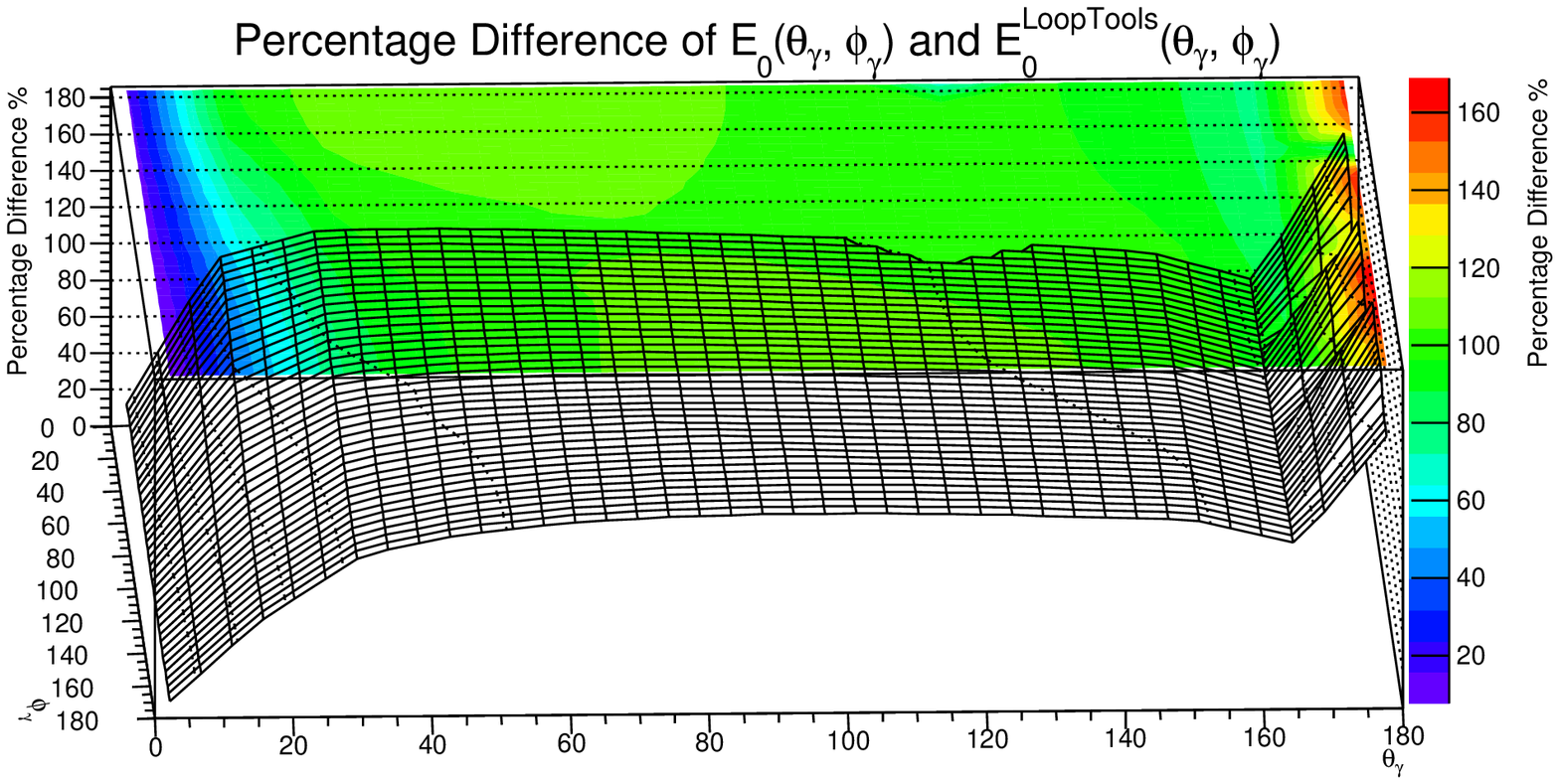}\\
		\caption{ Percentage difference of $E_0(\theta_\gamma,\phi_\gamma)$ and $E_0^{LoopTools}(\theta_\gamma,\phi_\gamma)$ for $\sqrt{s}=500\text{ GeV}$, $M_{V_1}=91\text{ GeV}$, $M_{V_1}=91\text{ GeV}$ and $\theta'_1=120^{\circ}$ }
	\end{center}
\end{figure}
Figs.~10 and 11, where $\theta'_1 = 105^{\circ},\; 120^{\circ}$, respectively, both have one region near the incoming $e^+$ where the two results differ by more than ~ 100\%. In both figures, they remain within a factor $\sim 2$ of one another.

\begin{figure}
	\begin{center}
		\includegraphics[scale=0.55]{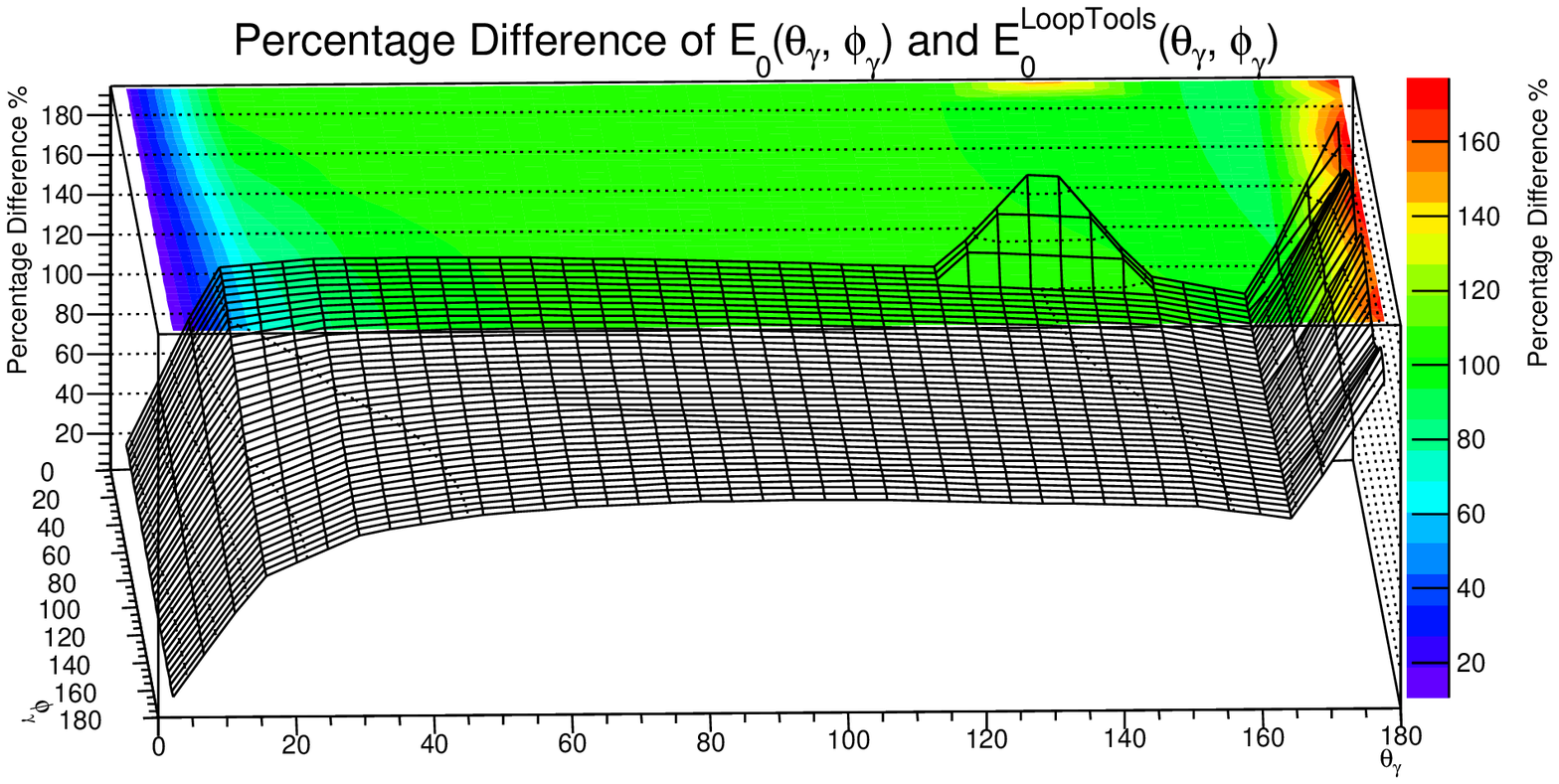}\\
		\caption{ Percentage difference of $E_0(\theta_\gamma,\phi_\gamma)$ and $E_0^{LoopTools}(\theta_\gamma,\phi_\gamma)$ for $\sqrt{s}=500\text{ GeV}$, $M_{V_1}=91\text{ GeV}$, $M_{V_1}=91\text{ GeV}$ and $\theta'_1=135^{\circ}$ }
	\end{center}
\end{figure}
In Fig.~12, where $\theta'_1 = 135^{\circ}$, there is a small region near the outgoing $\mu^-$ and a small region near the incoming $e^+$ where the two results differ by more than ~ 100\%. They remain within a factor $\sim 2$ of one another.

\begin{figure}
	\begin{center}
		\includegraphics[scale=0.55]{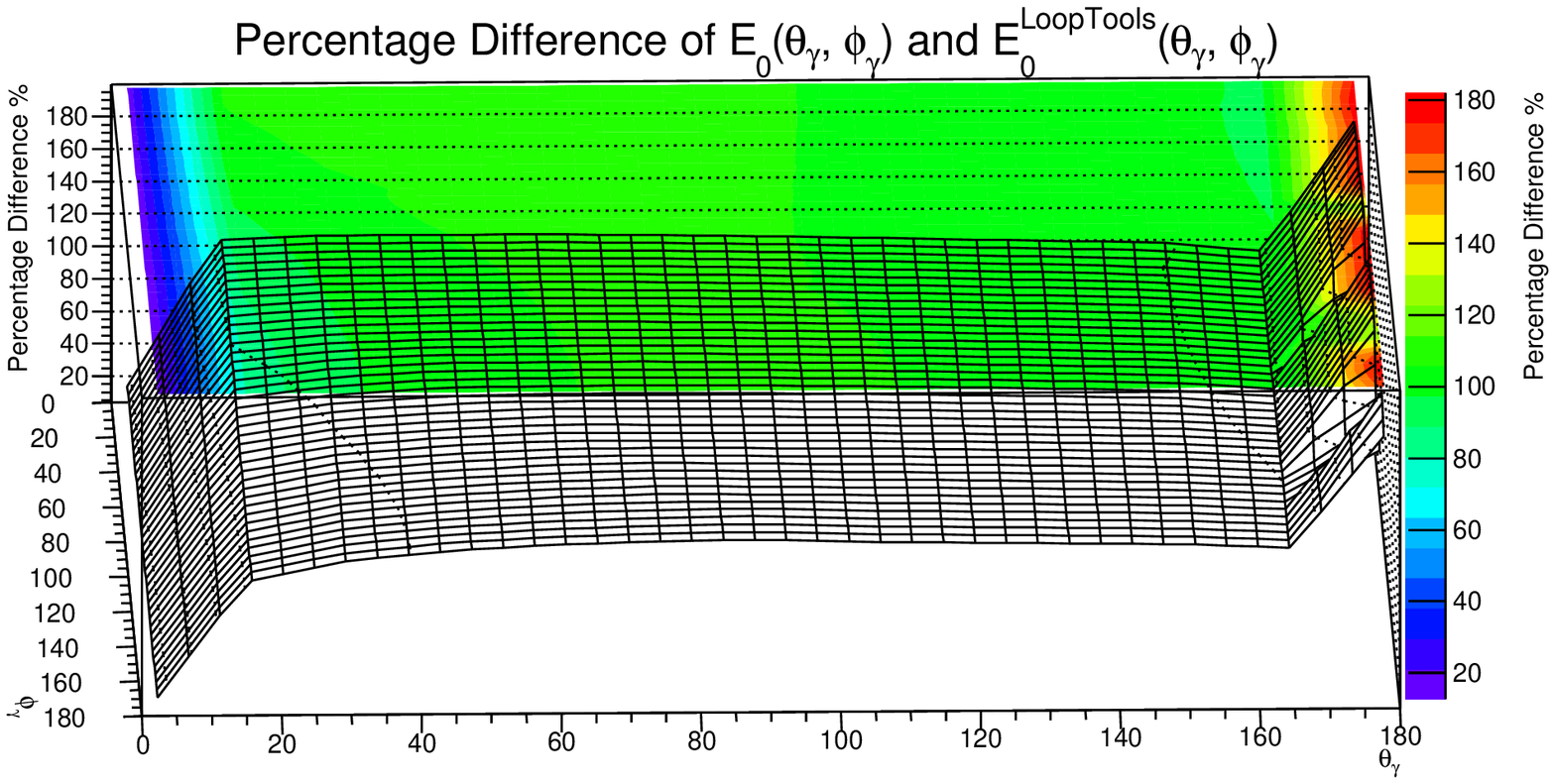}\\
		\caption{ Percentage difference of $E_0(\theta_\gamma,\phi_\gamma)$ and $E_0^{LoopTools}(\theta_\gamma,\phi_\gamma)$ for $\sqrt{s}=500\text{ GeV}$, $M_{V_1}=91\text{ GeV}$, $M_{V_1}=91\text{ GeV}$ and $\theta'_1=150^{\circ}$ }
	\end{center}
\end{figure}

\begin{figure}
	\begin{center}
		\includegraphics[scale=0.55]{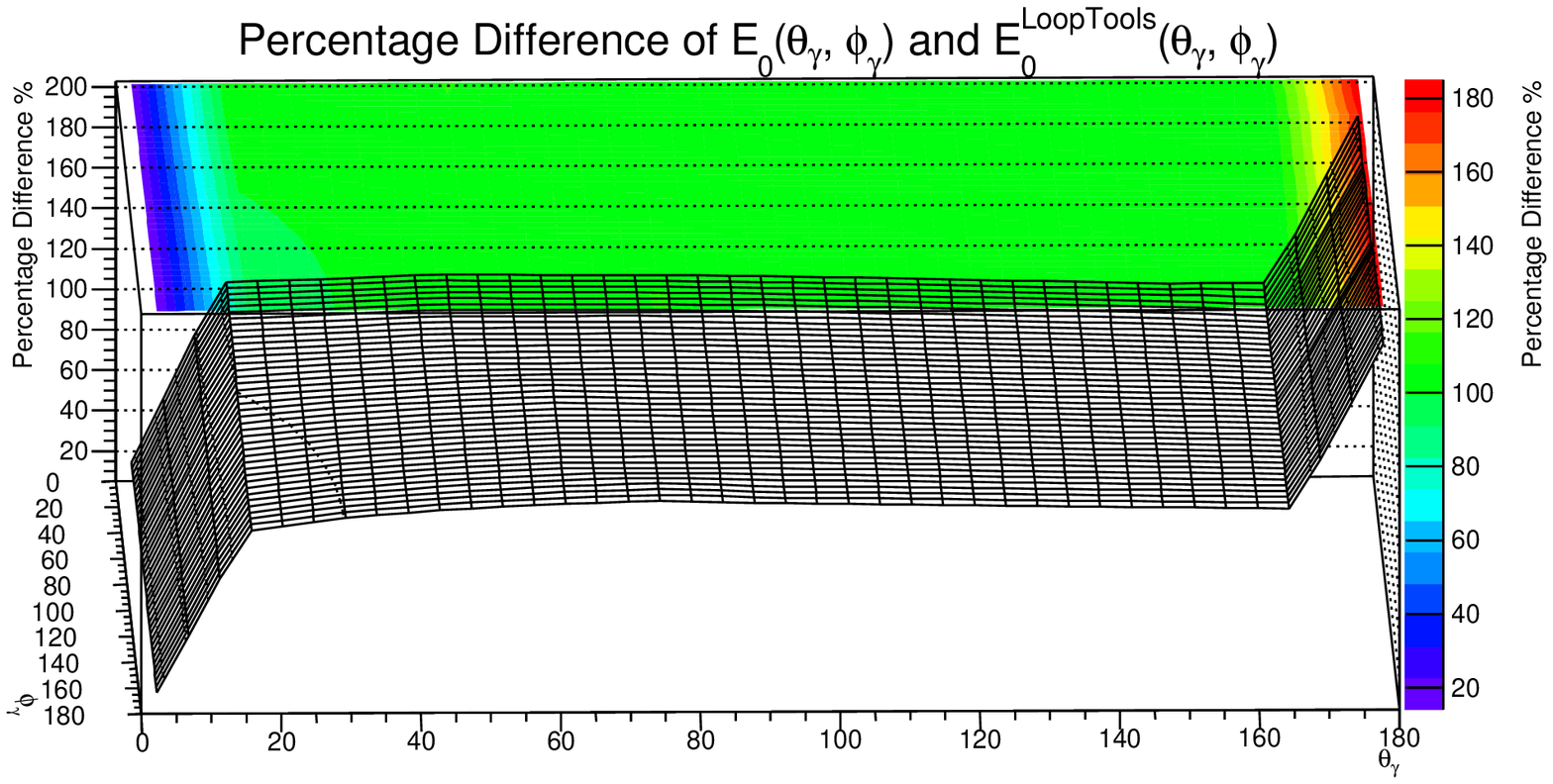}\\
		\caption{ Percentage difference of $E_0(\theta_\gamma,\phi_\gamma)$ and $E_0^{LoopTools}(\theta_\gamma,\phi_\gamma)$ for $\sqrt{s}=500\text{ GeV}$, $M_{V_1}=91\text{ GeV}$, $M_{V_1}=91\text{ GeV}$ and $\theta'_1=165^{\circ}$ }
	\end{center}
\end{figure}
Figs.~ 13 and 14, where $\theta'_1 = 150^{\circ},\; 165^{\circ}$, respectively, both have relatively smal regions near the incoming $e^+$ direction where the two results differ by more than ~ 100\%. In both figures, they remain within a factor $\sim 2$ of one another.

\begin{figure}
	\begin{center}
		\includegraphics[scale=0.55]{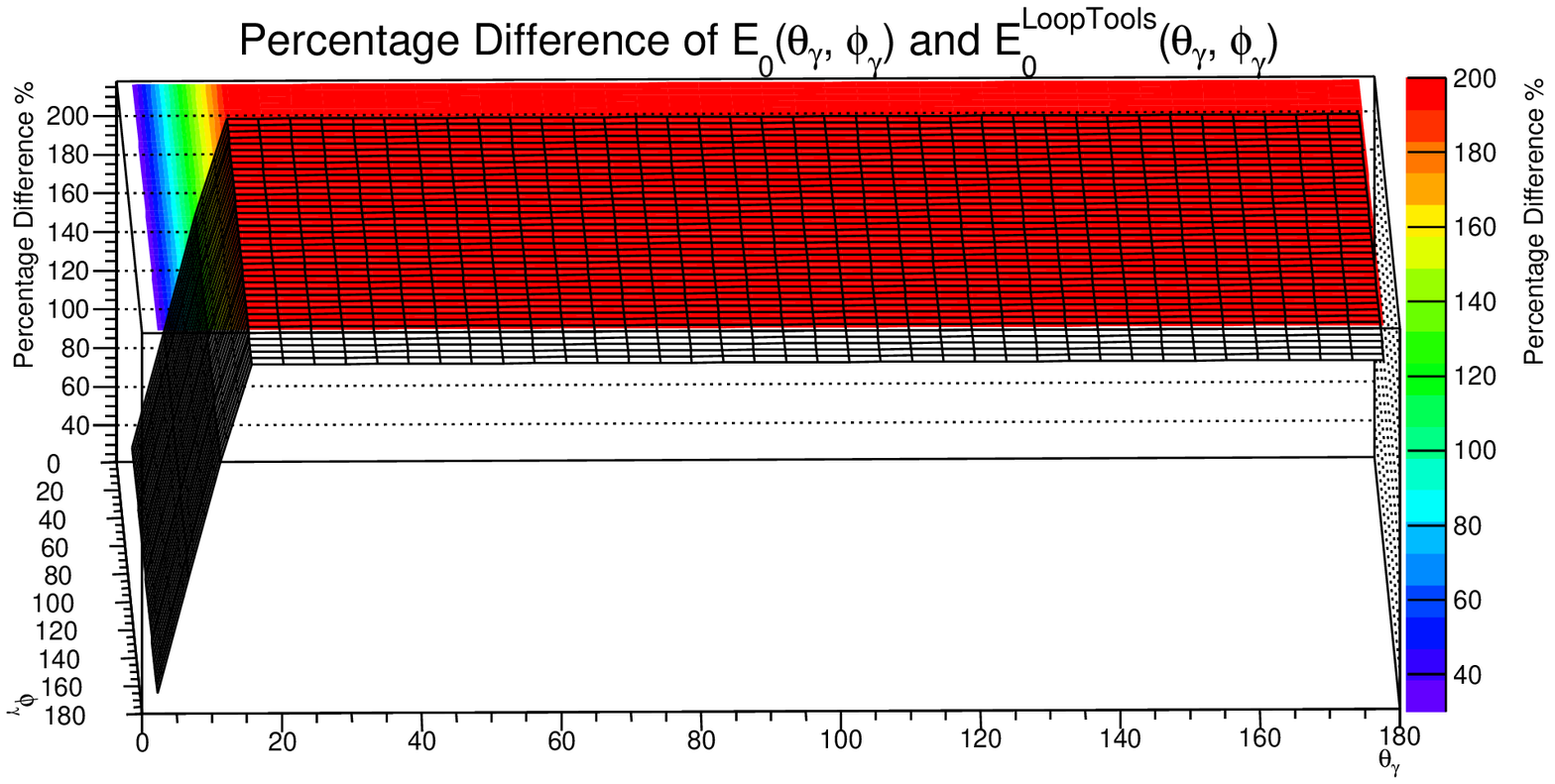}\\
		\caption{ Percentage difference of $E_0(\theta_\gamma,\phi_\gamma)$ and $E_0^{LoopTools}(\theta_\gamma,\phi_\gamma)$ for $\sqrt{s}=500\text{ GeV}$, $M_{V_1}=91\text{ GeV}$, $M_{V_1}=91\text{ GeV}$ and $\theta'_1=180^{\circ}$ }
	\end{center}
\end{figure}
\newpage
In Fig.~15, where where $\theta'_1 = 180^{\circ}$, there is a region near the incoming $e^-$ direction, which is $\theta_\gamma, \;\phi_\gamma) = (0^{\circ},\;0^\circ)$, independent of $\phi_\gamma$, where the two results are in good agreement. In the rest of the angular phase space, the two results differ by more than 160\%, though they remain within a factor of $\sim 2$ of one another.

As we see, the result from the magic spinor product method agrees with that from LoopTools in the overall sense, except for several regions. The percentage difference of our $E_0(\theta_\gamma,\phi_\gamma)\equiv E^{MS}_0(\theta_\gamma,\phi_\gamma) $ (MS denotes magic spinors) and $E_0^{\text{LoopTools}}(\theta_\gamma,\phi_\gamma)$ is insensitive with $\phi_\gamma$, but sensitive with $(\theta_\gamma,\;\theta'_1)$. When $\theta'_1>90^{\circ}$, our result mainly fits to that from LoopTools in the region $(0^{\circ}<\theta_\gamma<20^{\circ},\text{ } 0^{\circ}<\phi_\gamma<180^{\circ})$.
However, when $\theta'_1<75^{\circ}$, our result agrees greatly with that from LoopTools.

\section{Conclusion}
 In conclusion, we have developed a new numerical method to calculate the general five-point function, which is important for evaluating one-loop radiative corrections. Our method is developed from the magic spinor product approach in loop integrals proposed by one of us (BFLW) originally, which applied the "Chinese magic" spinor technique to simplify the loop integral so that the $E_0$ could be expressed more stably\footnote{The new $E_0\equiv E_{MS,0}$ based on magic spinor products in loop integrals is available from the authors upon request.} in terms of $n$-point one-loop integrals with $n\leq4$. The attendant $n$-point one-loop integrals $(n\leq4)$ can be calculated numerically by the package LoopTools. Theoretically, the magic spinor product method should provide more efficiency and numerical stability for the evaluation of the general five point function. By comparing the results obtained by our method with those directly obtained from LoopTools, we find that they agreed with each other in an overall sense. Such agreements are encouraging.
We are currently in the process of comparing with the approach in Ref.~\cite{collier}. The results of that comparison will appear elsewhere~\cite{elsewh}.
\section*{Acknowledgements}
We thank Prof. T. Hahn for helpful discussion. One of us (YL) thanks Prof. T. Steele for his kind hospitality at the University of Saskatchewan.

\bibliographystyle{unsrt}  


\begin{thebibliography}{1}

\bibitem{Denner} A. Denner, ``Techniques for the calculation of electroweak radiative corrections at the one-loop level and results for W-physics at LEP200", Fortschr. Phys. {\bf 41} (1993) 4.
\bibitem{PV} G. Passarino and M. Veltman, ``One-loop corrections for $e^+e^-$ annihilation into $\mu^+\mu^-$ in the Weinberg model", Nucl. Phys. B {\bf 160} (1979) 151.
\bibitem{MagicSpinor} B. F. L. Ward, ``Magic spinor product methods in loop integrals", Phys. Rev. D {\bf 83} (2011) 113014.
\bibitem{LT} Thomas Hahn, ``LoopTools 2.15 User's Guide", http://www.feynarts.de/looptools/LT215Guide.pdf.
\bibitem{vOV90} G. J. van Oldenborgh and J. A. M. Vermaseren, ``New algorithms for one-loop integrals", Z. Phys. C 46 (1990) 425.
	\bibitem{GPS} S. Jadach, B. F. L. Ward and Z. Was, ``Global positioning of spin GPS scheme for half-spin massive spinors", Eur. Phys. J. C {\bf 22} (2001) 423.
	
		\bibitem{JWZ2001prd} S. Jadach, B. F. L. Ward and Z. Was, ``Coherent exclusive exponentiation for precision Monte Carlo calculations", Phys. Rev. D {\bf 63} (2001) 113009.

	\bibitem{ChnMag} Z. Xu, D.-H. Zhang, and L. Change, ``Helicity amplitudes for multiple bremsstrahlung in massless non-abelian gauge theories", Nucl. Phys. B {\bf 291} (1987) 392.
\bibitem{KS} R. Kleiss, and W. J. Stirling, ``Spinor techniques for calculating $pp\to W^\pm/Z^0+\text{jets}$", Nucl. Phys. B {\bf 262} (1985) 235.

\bibitem{CALKUL} F. A. Berends, P. De Causmaecker, R. Gastmans,  R. Kleiss, W. Troost and T. T. Wu, (CALKUL Collaboration), ``Multiple bremsstrahlung in gauge theories at high energies: (III). Finite mass effects in collinear photon bremsstrahlung", Nucl. Phys. B {\bf 239} (1984) 382.

	\bibitem{tHscarlar} G. 't Hooft and M. Veltman, `` Scalar one-loop integrals", Nucl. Phys. B {\bf 153} (1979) 365.
		\bibitem{WJ} W. L. van Neerven and J. A. M. Vermaseren, ``Large loop integrals", Phys. Lett. B {\bf 137} (1984) 241.
	\bibitem{DD} A. Denner and S. Dittmaier, ``Reduction of one-loop tensor 5-point integrals", Nucl. Phys. B {\bf 658} (2003) 175.
\bibitem{DD1} A. Denner and S. Dittmaier, ``Reduction schemes for one-loop tensor integrals" Nucl. Phys. B {\bf 734} (2006) 62.

\bibitem{DG} D. Bardin and G. Passarino, ``The Standard Model in the Making",( Oxford University Press, Oxford,1999).
\bibitem{Consoli} M. Consoli, ``One-loop corrections to $e^+e^-\to e^+e^-$ in the Weinberg model", Nucl. Phys. B {\bf 160} (1979) 208.
\bibitem{collier} A. Denner, S. Dittmaier, and L. Hofer, "Collier: a fortran-based Complex One-Loop LIbrary in Extended Regularizations", Comput. Phys. Commun. {\bf 212} (2017) 220-238.
\bibitem{elsewh} Y. Liu and B.F.L. Ward, to appear.

\end{thebibliography}

\end{document}